\documentclass{appolb}
\usepackage{graphicx}

\begin{document}
\title{TDHF investigations of the U+U quasifission process%
\thanks{Presented at the XXXV Mazurian Lakes Conference on Physics, Piaski, Poland, September 3-9, 2017}%
}
\author{A.S. Umar
\address{Department of Physics and Astronomy, Vanderbilt University, Nashville, TN, 37235, USA \linebreak}\\
C. Simenel
\address{Department of Nuclear Physics, RSPE, The Australian National University, ACT 2601, Australia}
}

\maketitle
\begin{abstract}
The use of actinide collisions have been suggested as a way to produce neutron rich isotopes
of high Z nuclei. The collision dynamics of these reactions can be studied using unrestricted time-dependent
Hartree-Fock (TDHF) calculations. Here, we report on the recent  studies of quasifission for the
$^{238}$U+$^{238}$U system.
\end{abstract}

\section{Introduction}
The synthesis of superheavy elements (SHE) is one of the most exciting and challenging tasks
in nuclear physics~\cite{dullmann2015}. Nuclear density functional theories predict a
superheavy island of stability as a result of quantum mechanical shell closures~\cite{bender1999,nazarewicz2002, cwiok2005,pei2009a}.
Two approaches have been employed for the synthesis of SHE. In the so called \textit{cold fusion} reactions,
closed shell nuclei, such as
$^{208}$Pb (or $^{209}$Bi) are used as targets with projectile beams ranging from chromium
to zinc. These experiments were able to produce neutron-rich isotopes of elements
with $Z=107-112$~\cite{hofmann2002,munzenberg2015,morita2015}.
The choice of target and the low beam energies minimized the excitation energy for these reactions,
thus increasing the probability for evaporation residue formation by reducing other reaction processes
such as quasifission and fusion-fission.
As the limits of SHE formation have been reached in cold fusion reactions, an alternate approach, the
\textit{hot fusion} reactions, was embarked in search for higher $Z$ SHE.
In these reactions actinide targets were used with mainly a $^{48}$Ca beam.
Despite of the higher excitation energy,
isotopes of elements $Z=113-118$ were created~\cite{oganessian2015,roberto2015}.
In all these reactions the evaporation residue cross-section
is dramatically reduced due to the quasifission (QF) and fusion-fission (FF)
processes.
Quasifission occurs at a much shorter time-scale than fusion-fission~\cite{rietz2011}. 
Consequently, quasifission is the primary reaction
mechanism that limits the formation of superheavy nuclei. 
Studies have also shown a strong impact of the entrance channel characteristics, including deformation \cite{hinde1995,hinde1996,umar2006a,hinde2008,nishio2008,oberacker2014} and shell structure~\cite{simenel2012b} of the reactants.
The later stages of the dynamics are also impacted by the fissility of the total system~\cite{lin2012,rietz2013}, its neutron richness~\cite{hammerton2015}, and by shell effects in the exit channel~\cite{wakhle2014}. 
A number of models have been developed that describe the quasifission in terms of multi-nucleon transfer
process~\cite{adamian2003,zagrebaev2007,aritomo2009,zhao2016}.
Recently, time-dependent Hartree-Fock (TDHF) calculations have proven to be an excellent tool for studying QF dynamics, and
it particular mass-angle distributions and fragment
TKEs~\cite{wakhle2014,oberacker2014,hammerton2015,umar2015c,umar2016,prasad2016}.

Recently, actinide-actinide collisions have been suggested as a possible reaction mechanism to obtain neutron-rich
isotopes of high $Z$ nuclei as well as a possible means to search for SHE. The investigation of actinide-actinide collisions
have a rich history with various model studies, including the dinuclear system model (DNS)
model~\cite{feng2009a}, relativistic mean-field (RMF) and Skyrme HF
studies~\cite{gupta2007b}, reduced density-matrix formalism~\cite{sargsyan2009},
quantum molecular dynamics (QMD)~\cite{zhao2009} and improved quantum molecular dynamics
(ImQMD)~\cite{tian2008,zhao2016} calculations, as well as TDHF studies~\cite{cusson1980,golabek2009,kedziora2010,simenel2012}.
In this proceeding we present some of our recent results for the $^{238}$U+$^{238}$U system.

Frozen Hartree-Fock \cite{simenel2008,washiyama2008} calculations of potential 
between two $^{238}$U exhibit a barrier \cite{simenel2012}. 
However, this method neglects Pauli repulsion which increases with the mass of the nuclei \cite{simenel2017}, 
as well as effects of transfer which affect this potential dynamically \cite{umar2014a,vophuoc2016,liang2016,godbey2017}.
In addition, constrained HFB calculation of the fission path for the compound $^{238}$U+$^{238}$U system shows no fission barrier, 
indicating that the system is only expected to live for a short time.
Early TDHF calculations for this system (which used a plane of symmetry to save computational time) indeed indicated 
a maximum contact time of the order of $\sim4$~zs \cite{golabek2009}, in agreement with QMD simulations~\cite{tian2008}. 
Nevertheless, this time is long enough to enable significant transfer between the fragments, 
e.g., via an ``inverse quasifission'' mechanisms \cite{kedziora2010}.
The purpose of the present work is to study the characteristics of the fragments formed in $^{238}$U+$^{238}$U
using a TDHF code without spatial symmetries.

A brief introduction to the theoretical framework is provided in section~\ref{sec:TDHF},
followed by a presentation and discussion of the results in section~\ref{sec:results}.
Conclusions are drawn in section~\ref{sec:conclusions}.

\section{Formalism: TDHF and DC-TDHF} \label{sec:TDHF}
The TDHF theory allows us to study a large variety of phenomena
observed in low energy nuclear physics~\cite{simenel2012,negele1982}.
In particular, studies of nuclear reactions in the
vicinity of the Coulomb barrier, such as fusion, 
deep-inelastic reactions and transfer, 
and dynamics of quasifission~\cite{wakhle2014,oberacker2014,umar2015a,umar2015c,sekizawa2016} have been recently performed.

The TDHF equations for the single-particle wave functions
\begin{equation}
h(\{\phi_{\mu}\}) \ \phi_{\lambda} (r,t) = i \hbar \frac{\partial}{\partial t} \phi_{\lambda} (r,t)
            \ \ \ \ (\lambda = 1,...,A) \ ,
\label{eq:TDHF}
\end{equation}
can be derived from a time-dependent variational principle employing the same energy density functionals used in
nuclear structure calculations.
The study of dynamics contains no additional parameters.
The main approximation in TDHF is
that the many-body wave function is assumed to be a single time-dependent
Slater determinant at all times. It describes the time-evolution of the single-particle
wave functions in a mean-field corresponding to the dominant reaction channel.
During the past decade it has become numerically feasible to perform TDHF calculations on a
3D Cartesian grid without any symmetry restrictions
and with much more accurate numerical methods~\cite{umar2006c,maruhn2014}.
Furthermore, the quality of effective interactions has been substantially
improved~\cite{chabanat1998a,kluepfel2009,kortelainen2010}.

Recently, a new approach called the density constrained TDHF~\cite{umar2006b}
(DC-TDHF) was developed that facilitates
the  extraction of ion-ion interaction barriers as well as the excitation energies~\cite{umar2009a}
of the reaction products.
This approach is also used to calculate capture cross-sections and excitation energy of QF fragments, as
well as the dynamics of shape evolution during QF process~\cite{umar2010a,oberacker2014,umar2016,umar2015a}.

\section{Results}\label{sec:results}

In this proceeding, we focus on the reaction $^{238}$U+$^{238}$U. In our TDHF calculations
we used the Skyrme SLy4 energy density functional~\cite{chabanat1998a}
including all of the relevant time-odd terms in the mean-field Hamiltonian.
\begin{figure}[!htb]
\centering
\includegraphics*[width=10cm]{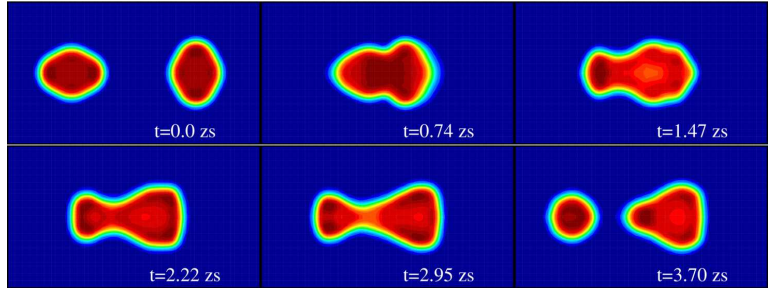}
\caption{\protect Quasifission in a central collision of $^{238}$U+$^{238}$U
		at $E_{\mathrm{c.m.}}=1350$~MeV.
                The initial orientation of the deformed nuclei is "tip-side".
		Shown is a contour plot of the time evolution of the mass density. Time
		increases from left to right and top to bottom. The heavy fragment
                in the exit channel has average charge $Z\simeq124$ and mass $A\simeq325$.}
\label{fig1}
\end{figure}
The reason for using SLy4
was due to the availability of the pairing force parameters for this force, suitable for our code.
To describe these reactions with a high degree of accuracy, the shapes of the individual
nuclei must be correctly reproduced by the mean-field theory. In some cases, it is necessary to
include BCS pairing which increases the number of single-particle levels that must
be taken into account by about $50\%$. It turns out that including
BCS pairing for the neutrons in $^{238}$U (using fixed partial occupations) in the
TDHF runs produces high quality axially symmetric nuclear shapes with prolate quadrupole
and hexadecapole deformations.

Numerically, we proceed as follows: First we generate very well-converged static
HF-BCS wave functions for the two nuclei on a 3D Cartesian grid. For most runs, we used a
lattice with $70*40*40$ grid points in $x,y,z$ directions, with an initial separation of the
two nuclei of $R=34$~fm. To test the numerical accuracy, we also carried out one
run on a substantially larger lattice with $90*60*60$ grid points in $x,y,z$ directions,
with an initial separation of the two nuclei of $R=40$~fm. 
Note that, although Coulomb reorientation of the deformed nuclei could in principle happen before this initial distance, 
it is only expected to be significant with a light deformed nucleus on a heavy collision partner \cite{simenel2004} and 
can therefore be neglected here. 

In the second
step, we apply a boost operator to the single-particle wave functions. The time-propagation
is carried out using a Taylor series expansion (up to orders $10-12$) of the
unitary mean-field propagator, with a time step $\Delta t = 0.4$~fm/c.
\begin{figure}[!htb]
\centering
\includegraphics*[width=0.5\linewidth]{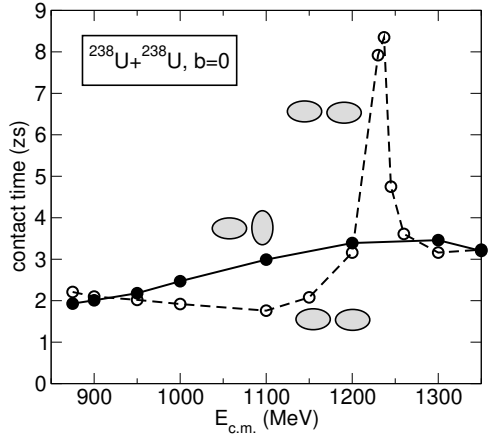}
\caption{\protect Nuclear contact time for central collisions of $^{238}$U+$^{238}$U
                as a function of center-of-mass energy.
                Results are shown for two initial orientations of the deformed nuclei:
                "tip-side" and "tip-tip".\label{fig2}}
\end{figure}
For very heavy systems such as $^{238}$U+$^{238}$U, the TDHF calculations of QF
require very long CPU times (due to long reaction time): a single TDHF run at fixed $E_\mathrm{c.m.}$ energy
and zero impact parameter takes about 2 weeks of CPU time
on a 16-processor LINUX workstation. A total CPU time of about 6 months was required
for all of the calculations presented in this contribution.
In Fig.~\ref{fig1} we show an example contour plot of the mass density in the $x$-$z$ plane as a function of time
for the central collision of the $^{238}$U+$^{238}$U system at $E_{\mathrm{c.m.}}=1350$~MeV and initial tip-side
orientation.

We define the contact time for QF as the time interval between the time $t_1$
when the two nuclear surfaces
(defined as isodensities with half the saturation density $\rho_0/2=0.08$~fm$^{-3}$)
first merge into a single surface and
the time $t_2$ when the surface splits up again.
Figure~\ref{fig2} shows the contact time as a function
of center-of-mass energy for central collisions of the $^{238}$U+$^{238}$U system, for both
tip-tip and tip-side orientations.
We observe that the tip-side collisions generally have a longer contact time, which is expected since the
two nuclei can reach a more compact initial shape, which increases the lifetime of the composite system.
This trend seems to break down at high enough energy, e.g. around $E_{\mathrm{c.m.}}=1230$~MeV, at which point
tip-tip collisions also start to have significant density overlap.
However, at higher energies the contact time is again reduced. This unusual increase of contact time
for tip-tip collisions around $1237$~MeV can be attributed to the energy dependence of the dynamically changing shell effects
and the breaking of the initial symmetry by the code. The TDHF program does not contain any symmetries and therefore
for processes involving long-time scales it can break the symmetry if it is energetically favorable~\cite{umar2010c}.
\begin{figure}[!hbt]
  \begin{minipage}{0.482\linewidth}
    \centering
    \includegraphics*[width=.9\linewidth]{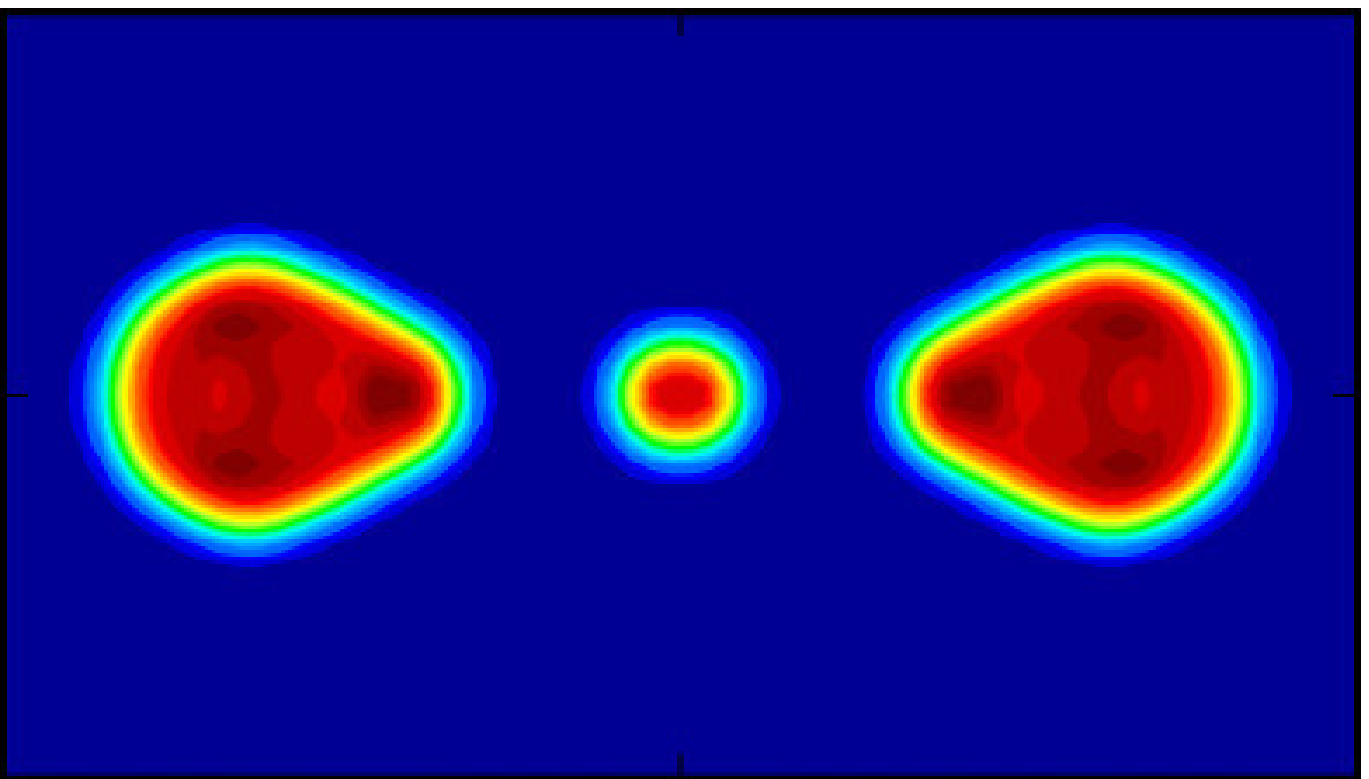}
    \caption{Exit channel mass density distribution for the central tip-tip collision of
                $^{238}$U+$^{238}$U at $E_{\mathrm{c.m.}}=875$~MeV.
                Fragment in the middle has charge $Z\simeq7$ and mass $A\simeq18$.\label{fig3}}
    \vspace{3ex}
  \end{minipage}
  \begin{minipage}{0.482\linewidth}
    \centering
    \includegraphics*[width=.9\linewidth]{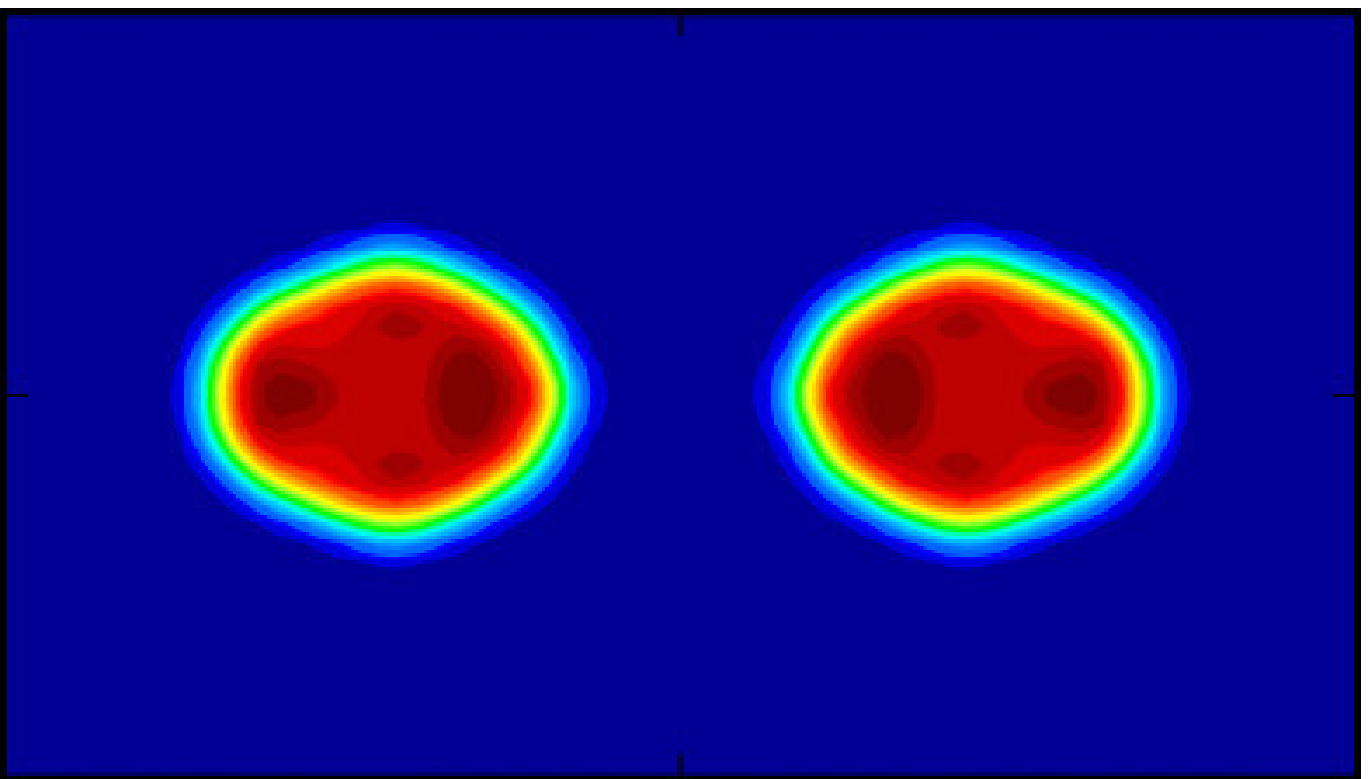}
    \caption{Exit channel mass density distribution for central tip-tip collision of
                $^{238}$U+$^{238}$U at $E_{\mathrm{c.m.}}=1100$~MeV.
                Two equal fragments of $^{238}$U are observed.\label{fig4}}
    \vspace{3ex}
  \end{minipage}
  \begin{minipage}{0.482\linewidth}
    \centering
    \includegraphics*[width=.9\linewidth]{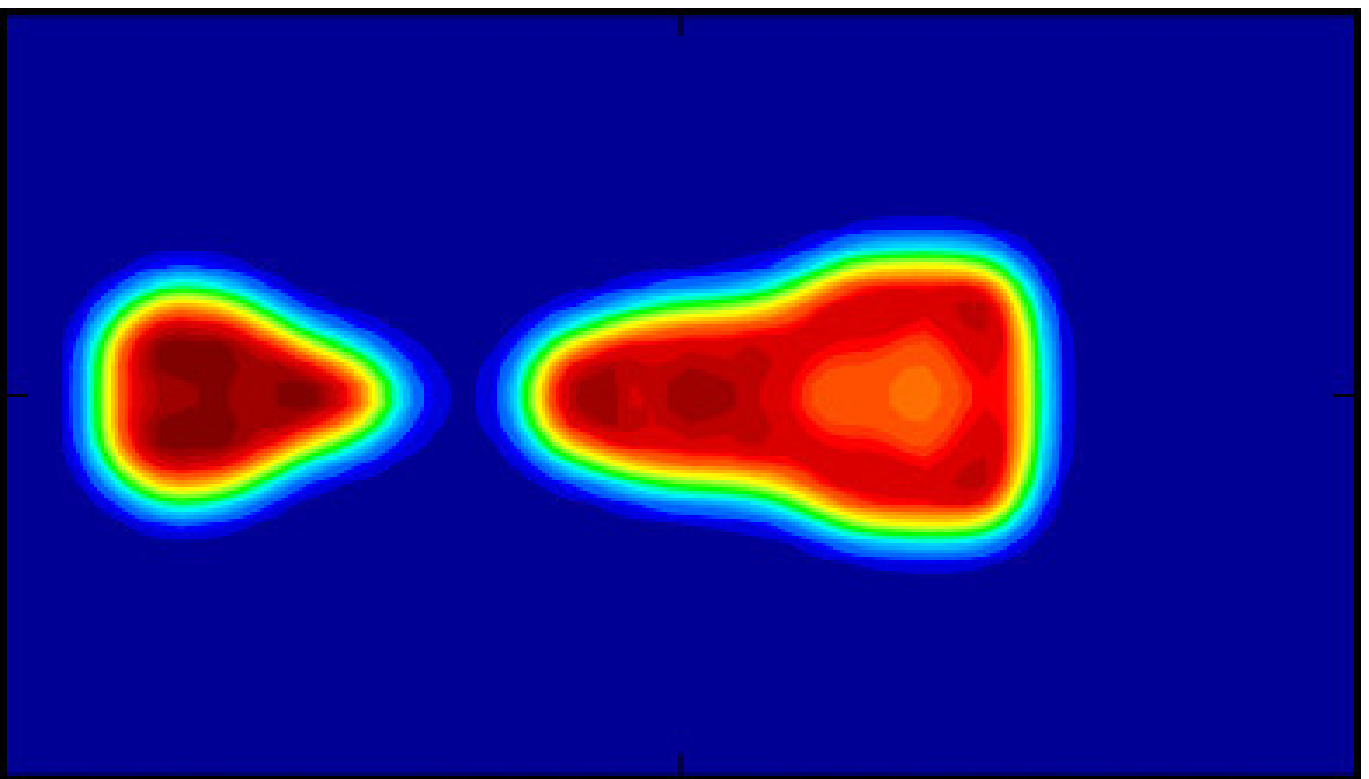}
    \caption{Exit channel  mass density distribution for central tip-tip collision of
                $^{238}$U+$^{238}$U at $E_{\mathrm{c.m.}}=1237$~MeV.
                The heavy fragment has $Z\simeq134$ and $A\simeq353$.\label{fig5}}
  \end{minipage}
  \hspace{0.1in}
  \begin{minipage}{0.482\linewidth}
    \centering
    \includegraphics*[width=.9\linewidth]{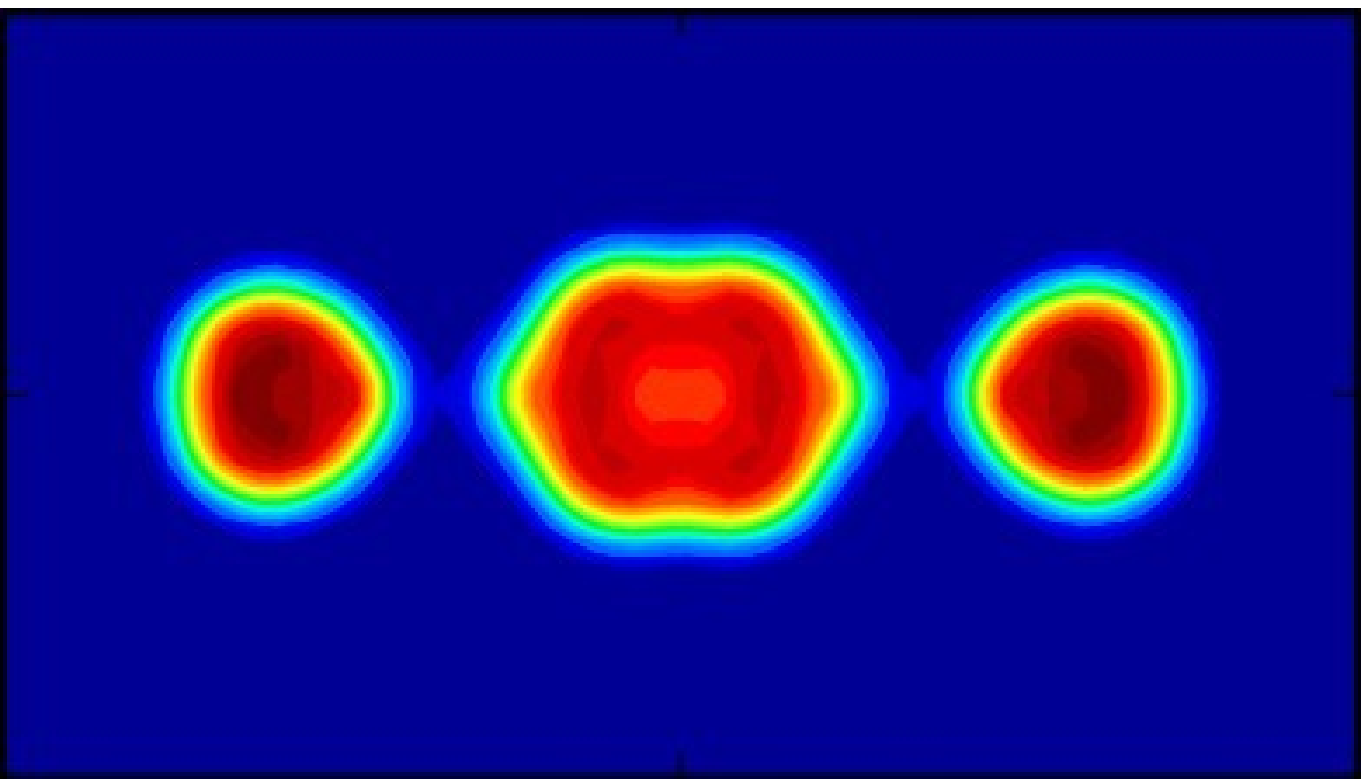}
    \caption{Exit channel mass density distribution for central tip-tip collision of
                $^{238}$U+$^{238}$U at $E_{\mathrm{c.m.}}=1300$~MeV.
                Three fragments are observed with middle fragment $Z\simeq103$ and $A\simeq274$.\label{fig6}}
  \end{minipage}
\end{figure}
For the lower energy central tip-tip collisions we sometimes get ternary quasifission with a light neutron rich fragment in the middle and two large fragments on each side. This was also seen in earlier
TDHF calculations with a plane of symmetry~\cite{golabek2009}.
This behavior changes as we increase the beam energy.
In Figs.~\ref{fig3}-\ref{fig6} we see samples of the energy dependence of the central tip-tip collisions of $^{238}$U+$^{238}$U.
Figure~\ref{fig3} corresponds to the lowest energy collision at $E_{\mathrm{c.m.}}=875$~MeV.
In this case exit
channel contains three fragments, with a small central fragment with charge $Z\simeq7$ and mass $A\simeq18$.
The light fragments left at the center for these lower energy collisions can be attributed to physics of neck
fragmentation~\cite{golabek2009}.
At the higher energy of $E_{\mathrm{c.m.}}=1100$~MeV we again end up with essentially two
excited $^{238}$U nuclei in the exit channel as shown in Fig.~\ref{fig4}.
The two fragment exit channel continues until the collision energy of $E_{\mathrm{c.m.}}=1237$~MeV. At this
energy the contact time peaks as shown in Fig.~\ref{fig2}. The heavy fragment in the exit channel has
charge $Z\simeq134$ and mass $A\simeq353$, as shown in Fig.~\ref{fig5}.
\begin{figure}[!hbt]
\includegraphics*[width=4cm]{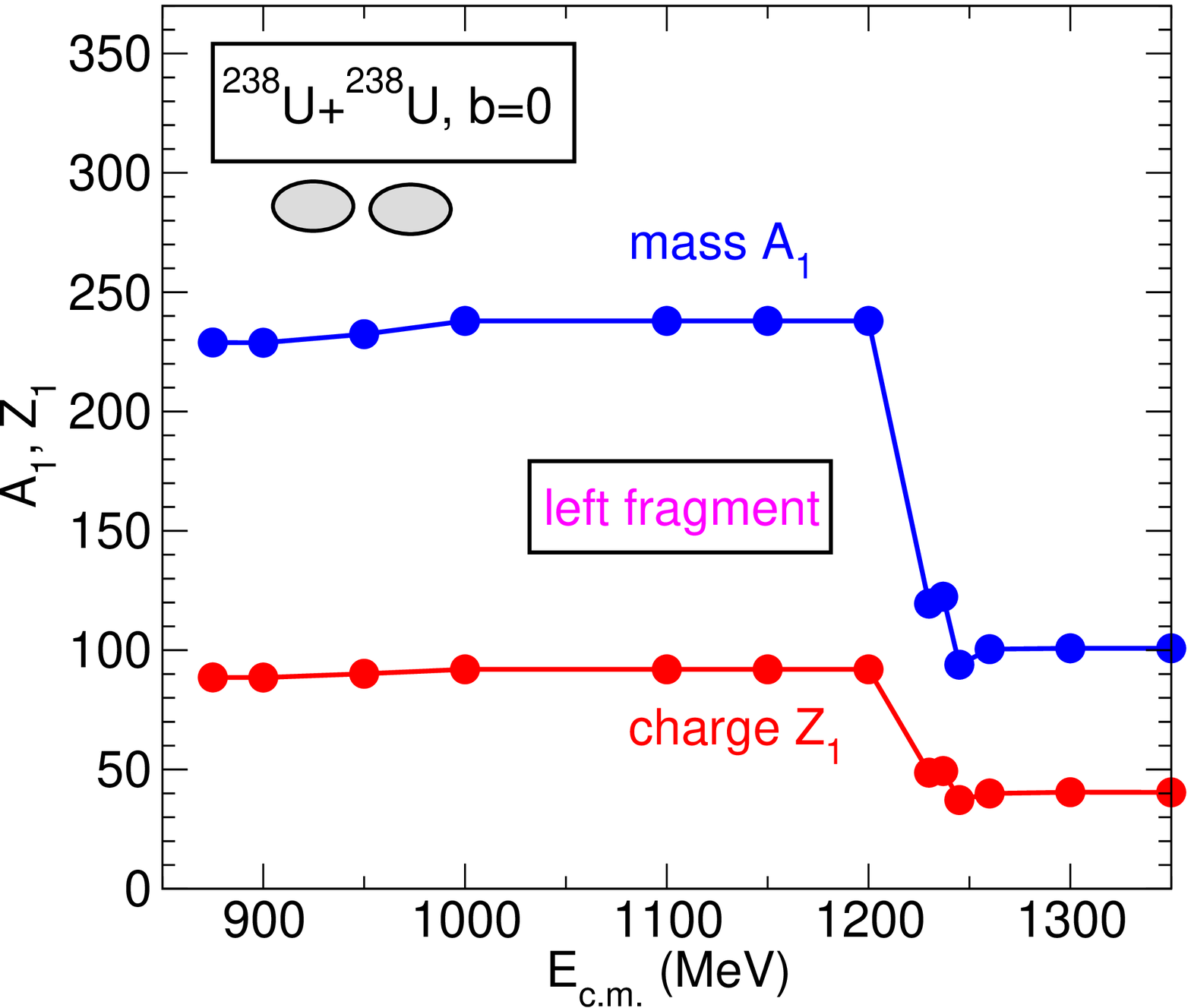}
\includegraphics*[width=4cm]{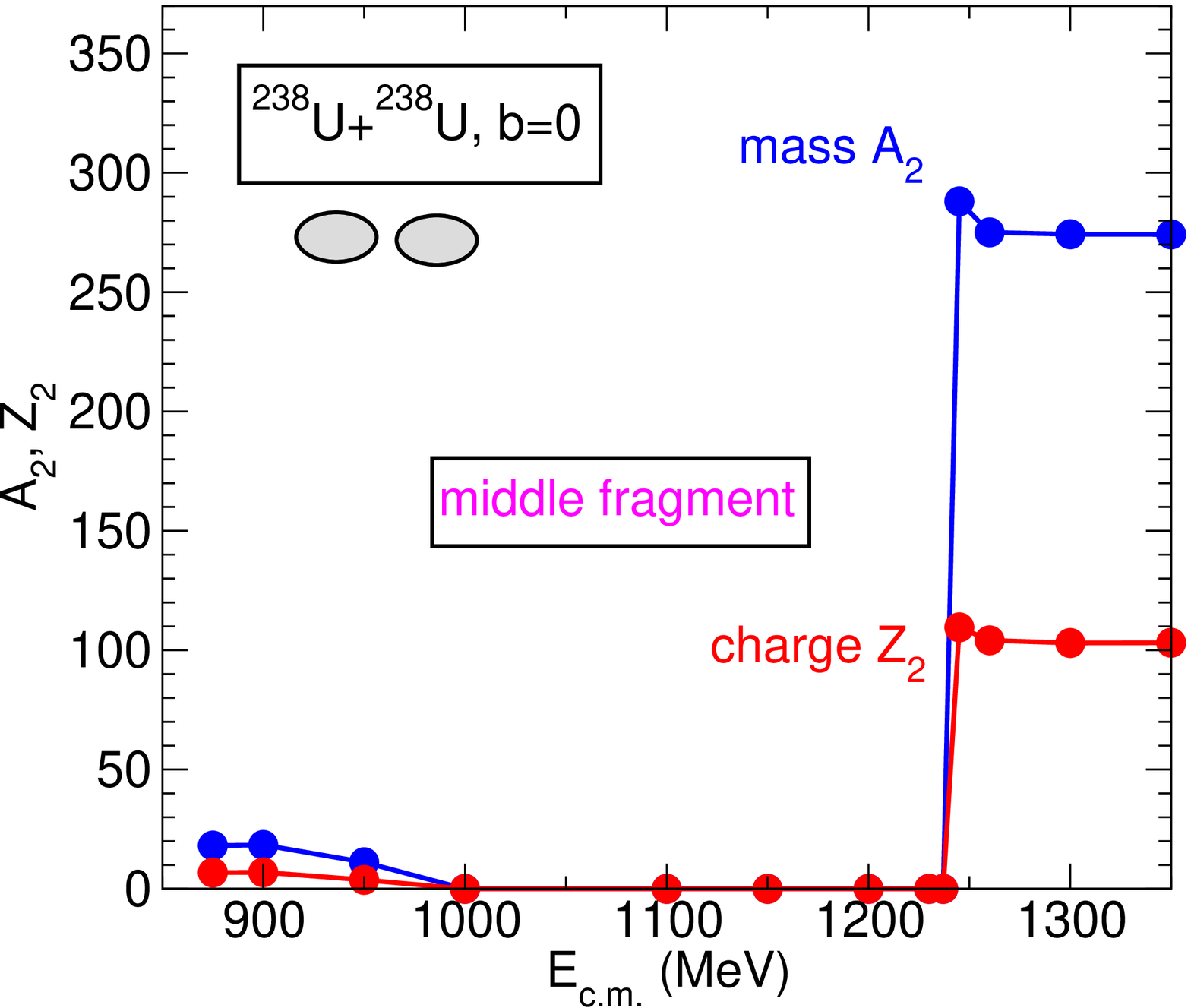}
\includegraphics*[width=4cm]{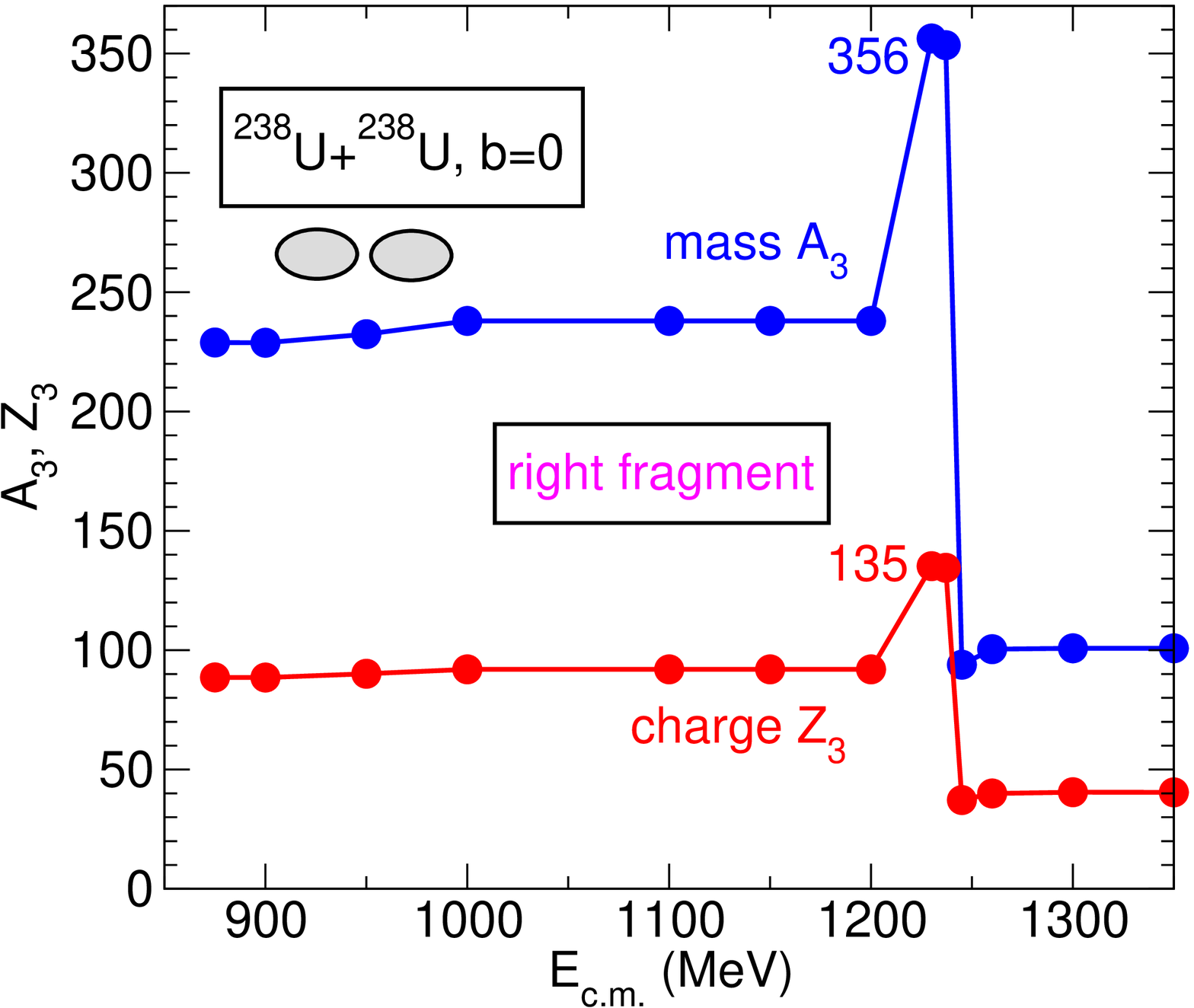}
\caption{\protect Mass and charge of the left fragment, middle fragment (if any)
and right fragment as a function of $E_\mathrm{c.m.}$ for central collisions
of $^{238}$U+$^{238}$U. The initial orientation of the deformed nuclei is "tip-tip".}
\label{fig7}
\end{figure}
We have done three calculations in the energy interval $E_{\mathrm{c.m.}}=1230-1243$~MeV to confirm the
unexpected peak in the contact time for tip-tip collisions, shown in Fig.~\ref{fig2}.
At the highest energy of $E_{\mathrm{c.m.}}=1300$~MeV we again have ternary quasifission but in this
case the middle fragment is the heaviest one with charge $Z\simeq103$ and mass $A\simeq274$, as shown in Fig.~\ref{fig6}.
These results are also in good agreement with the earlier findings from TDHF with a plane of symmetry~\cite{golabek2009}.
The summary of all of our results for the central tip-tip collisions are shown in Fig.~\ref{fig7}
as a function of $E_\mathrm{c.m.}$.

\begin{figure}[!htb]
\begin{minipage}{0.45\linewidth}
\centering
\includegraphics*[width=\linewidth]{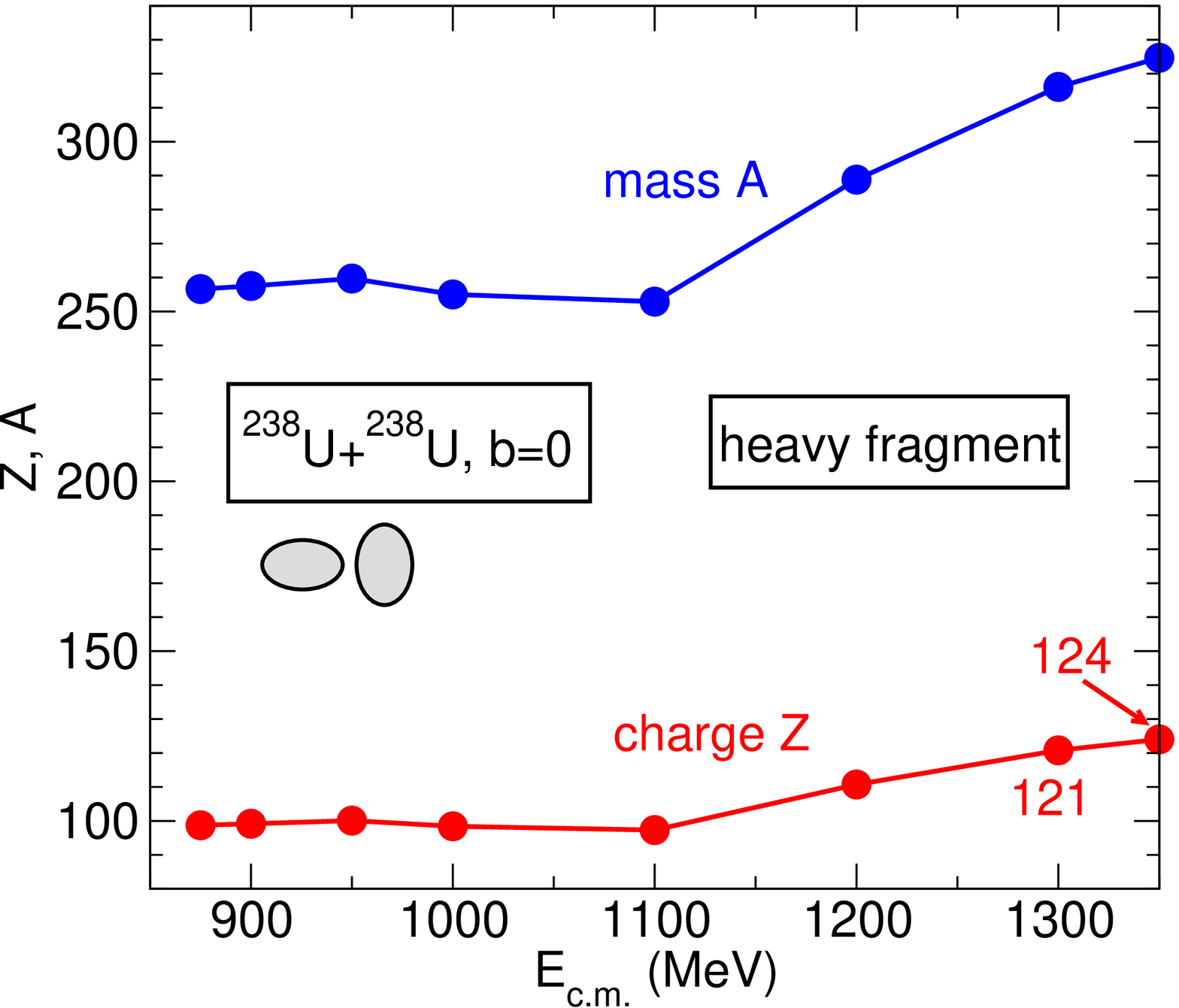}
\caption{\protect Mass and charge of the heavy fragment
as a function of $E_\mathrm{c.m.}$ for central collisions
of $^{238}$U+$^{238}$U. The initial orientation of the deformed nuclei is "tip-side".\label{fig8}}
\end{minipage}%
\hspace{0.2in}
\begin{minipage}{0.45\linewidth}
\centering
 \includegraphics*[width=\linewidth]{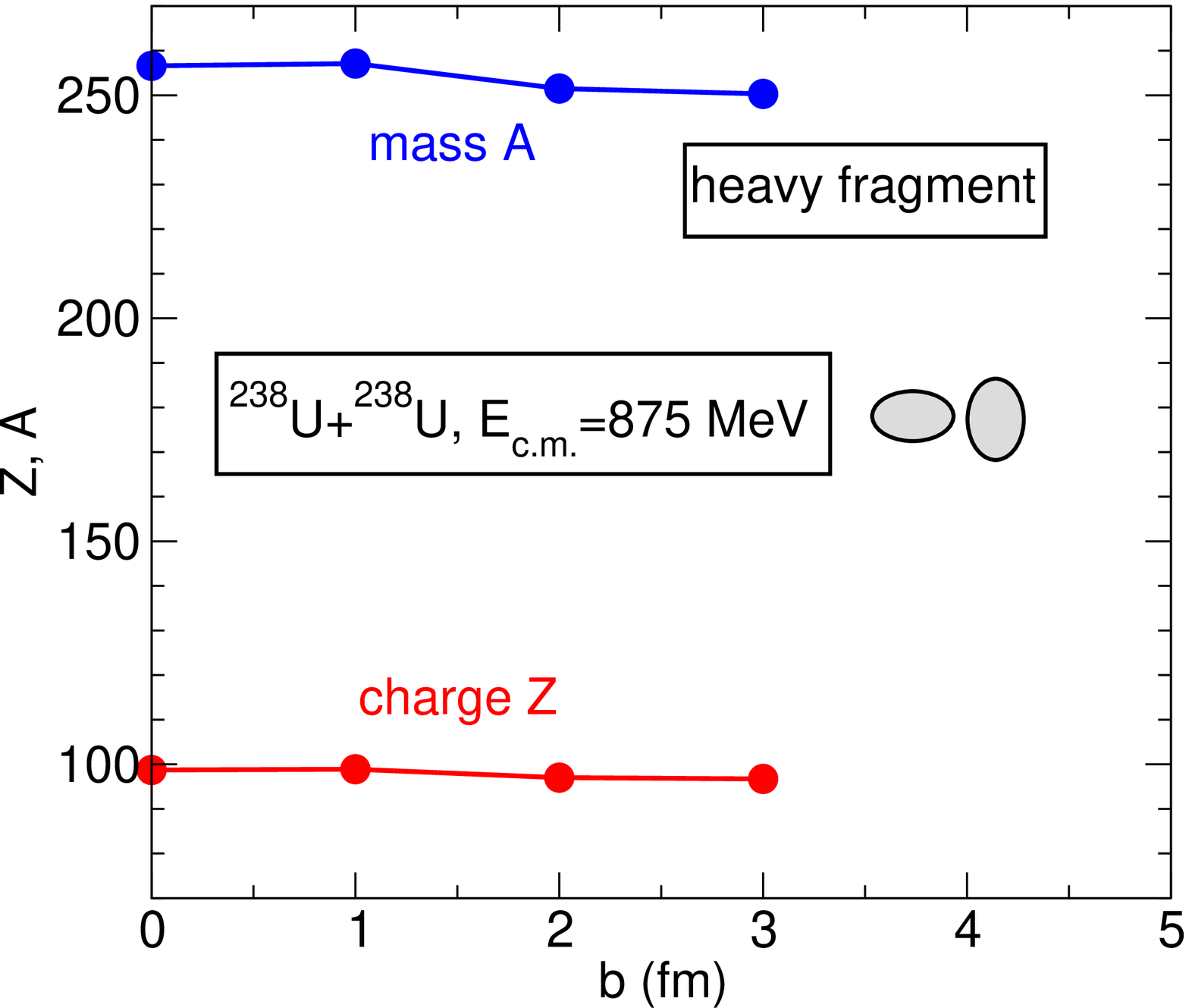}
\caption{\protect Mass and charge of the heavy fragment
as a function of impact parameter $b$ for $^{238}$U+$^{238}$U collisions at
$E_\mathrm{c.m.}=875$~MeV. The initial orientation of the deformed nuclei is "tip-side".\label{fig9}}
\end{minipage}
\end{figure}

We next discus the energy dependence of the central tip-side collisions. As we have seen in Fig.~\ref{fig2}
the contact time varies slightly over the entire energy range studied here.
From Fig.~\ref{fig8} we observe that the mass and charge transfer to the heavy fragment are
roughly proportional to the nuclear contact time.
Until about $E_{\mathrm{c.m.}}=1100$~MeV the heavy fragment stays in the range $Z=98-101$ and $A=252-260$.
At higher energy there is a steady increase in the charge and mass of the heavy fragment.
At the highest energy of $E_{\mathrm{c.m.}}=1350$~MeV
we observe two fragments in the exit channel with the heavy fragment having charge $Z=124$ and mass $A=325$.
The corresponding mass density plot was shown in the last frame of Fig.~\ref{fig1}.
We have also examined the impact parameter dependence for the lowest energy $E_\mathrm{c.m.}=875$~MeV
as shown in Fig.~\ref{fig9}. At this energy the trend is very smooth up to an impact parameter of
$b=3$~fm. After that we essentially get two excited $^{238}$U nuclei in the exit channel.

\begin{figure}[!htb]
\centering
\includegraphics*[width=0.5\linewidth]{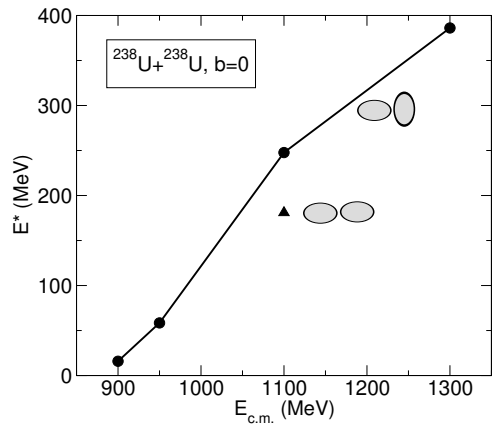}
\caption{\protect Excitation energy of the heavy-fragment for the central collisions of $^{238}$U+$^{238}$U
                as a function of center-of-mass energy.
                Results are shown for two initial orientations of the deformed nuclei:
                "tip-side" and "tip-tip".\label{fig10}}
\end{figure}

One of the questions to be asked in the light of these results is the excitation energy of the outgoing
fragments since this would indicate the survival probability of the fragments to fission.
We have
calculated few excitation energies for the tip-side collision and one for the tip-tip collision.
We are unable to compute the excitation energy in the case of ternary quasifission due to technical
issues. In Fig.~\ref{fig10} we show these excitation energy of the heavy fragment. While, at lower beam energies the
excitation energy is relatively small there is almost a linear increase of excitation energy with increasing
c.m. energy. In cases were high-$Z$ and high-$A$ fragments are obtained the excitation energy is in the
hundreds of MeV, thus challenging the likelihood of survival for these fragments.

\section{Conclusions}\label{sec:conclusions}
We have performed TDHF calculations for the $^{238}$U+$^{238}$U system as a function of c.m. energy for
central collisions and in one case as a function of impact parameter. At lower energies, that are in
the range of current experimental studies, TDHF results do not show high-$Z$ and high-$A$ fragments but
rather the possibility of producing neutron rich lighter elements. At higher energies some high-$Z$ and high-$A$
primary fragments are observed. However, the large excitation energies associated with these fragments makes 
their survival probabilities very low. On the other hand formation of neutron rich isotopes of high-$Z$ nuclei
may be possible. This particular study has limited results for impact parameter dependence of these
collisions as well as collisions involving arbitrary orientations of the two deformed $^{238}$U nuclei.
Nevertheless, this study indicates that  these collisions may serve as a means to produce neutron rich isotopes.
Further calculations will be done to examine this point.
Beyond mean-field techniques~\cite{balian1992,lacroix2014}
could also be applied to obtain fragment mass and kinetic energy distributions as it was recently done~\cite{simenel2012,tanimura2017,ayik2017}.
In addition, it would be interesting to investigate the distribution of cold fragments 
by coupling TDHF with a statistical decay code~\cite{sekizawa2017,umar2017}.

\section*{Acknowledgments}
This work has been supported by the U.S. Department of Energy under grant No.
DE-SC0013847 with Vanderbilt University and by the
Australian Research Council Grant Nos. FL110100098, FT120100760, DP160101254.


\begin{thebibliography}{99}

\bibitem{dullmann2015}
{Christoph E. D\"ullmann}, {Rolf--Dietmar Herzberg}, {Witold Nazarewicz}, {Yuri
  Oganessian},
 Nucl. Phys. A
{\bf  944}, 1 (2015),  {S}pecial {I}ssue on {S}uperheavy
  {E}lements.

\bibitem{bender1999}
M.~Bender, K.~Rutz, {P.--G.} Reinhard, J.~A. Maruhn, W.~Greiner,
 Phys. Rev. C
{\bf  60}, 034304 (1999).

\bibitem{nazarewicz2002}
W.~Nazarewicz, M.~Bender, S.~\'{C}wiok, P.H. Heenen, A.T. Kruppa, {P.--G.}
  Reinhard, T.~Vertse,
 Nucl. Phys. A
{\bf  701}, 165 (2002).

\bibitem{cwiok2005}
S.~\'{C}wiok, P.-H. Heenen, W.~Nazarewicz,
 Nature
{\bf  433}, 705 (2005).

\bibitem{pei2009a}
J.~C. Pei, W.~Nazarewicz, J.~A. Sheikh, A.~K. Kerman,
 Phys. Rev. Lett.
{\bf  102}, 192501 (2009).

\bibitem{hofmann2002}
S.~Hofmann, F.~P. He\ss{}berger, D.~Ackermann, G.~M\"unzenberg, S.~Antalic,
  P.~Cagarda, B.~Kindler, J.~Kojouharova, M.~Leino, B.~Lommel, R.~Mann, A.~G.
  Popeko, S.~Reshitko, S.~\'Saro, J.~Uusitalo, A.~V. Yeremin,
 Eur. Phys. J. A
{\bf  14}, 147 (2002).

\bibitem{munzenberg2015}
G.~M\"unzenberg, K.~Morita,
 Nucl. Phys. A
{\bf  944}, 3 (2015).

\bibitem{morita2015}
Kosuke Morita,
 Nucl. Phys. A
{\bf  944}, 30 (2015).

\bibitem{oganessian2015}
{Yu. Ts. Oganessian}, V.~K. Utyonkov,
 Nucl. Phys. A
{\bf  944}, 62 (2015).

\bibitem{roberto2015}
J.~B. Roberto, C.~W. Alexander, R.~A. Boll, J.~D. Burns, J.~G. Ezold, L.~K.
  Felker, S.~L. Hogle, K.~P. Rykaczewski,
 Nucl. Phys. A
{\bf  944}, 99 (2015).

\bibitem{rietz2011}
R.~{du Rietz}, D.~J. Hinde, M.~Dasgupta, R.~G. Thomas, L.~R. Gasques, M.~Evers,
  N.~Lobanov, A.~Wakhle,
 Phys. Rev. Lett.
{\bf  106}, 052701 (2011).

\bibitem{hinde1995}
D.~J. Hinde, M.~Dasgupta, J.~R. Leigh, J.~P. Lestone, J.~C. Mein, C.~R. Morton,
  J.~O. Newton, H.~Timmers,
 Phys. Rev. Lett.
{\bf  74}, 1295 (1995).

\bibitem{hinde1996}
D.~J. Hinde, M.~Dasgupta, J.~R. Leigh, J.~C. Mein, C.~R. Morton, J.~O. Newton,
  H.~Timmers,
 Phys. Rev. C
{\bf  53}, 1290 (1996).

\bibitem{umar2006a}
A.~S. Umar, V.~E. Oberacker,
 Phys. Rev. C
{\bf  74}, 061601 (2006).

\bibitem{hinde2008}
D.~J. Hinde, R.~G. Thomas, R.~du~Rietz, A.~Diaz-Torres, M.~Dasgupta, M.~L.
  Brown, M.~Evers, L.~R. Gasques, R.~Rafiei, M.~D. Rodriguez,
 Phys. Rev. Lett.
{\bf  100}, 202701 (2008).

\bibitem{nishio2008}
K.~Nishio, H.~Ikezoe, S.~Mitsuoka, I.~Nishinaka, Y.~Nagame, Y.~Watanabe,
  T.~Ohtsuki, K.~Hirose, S.~Hofmann,
 Phys. Rev. C
{\bf  77}, 064607 (2008).

\bibitem{oberacker2014}
V.~E. Oberacker, A.~S. Umar, C.~Simenel,
 Phys. Rev. C
{\bf  90}, 054605 (2014).

\bibitem{simenel2012b}
C.~Simenel, D.~J. Hinde, R.~{du Rietz}, M.~Dasgupta, M.~Evers, C.~J. Lin, D.~H.
  Luong, A.~Wakhle,
 Phys. Lett. B
{\bf  710}, 607 (2012).

\bibitem{lin2012}
C.~J. Lin, R.~du~Rietz, D.~J. Hinde, M.~Dasgupta, R.~G. Thomas, M.~L. Brown,
  M.~Evers, L.~R. Gasques, M.~D. Rodriguez,
 Phys. Rev. C
{\bf  85}, 014611 (2012).

\bibitem{rietz2013}
R.~du~Rietz, E.~Williams, D.~J. Hinde, M.~Dasgupta, M.~Evers, C.~J. Lin, D.~H.
  Luong, C.~Simenel, A.~Wakhle,
 Phys. Rev. C
{\bf  88}, 054618 (2013).

\bibitem{hammerton2015}
K.~Hammerton, Z.~Kohley, D.~J. Hinde, M.~Dasgupta, A.~Wakhle, E.~Williams,
  V.~E. Oberacker, A.~S. Umar, I.~P. Carter, K.~J. Cook, J.~Greene, D.~Y.
  Jeung, D.~H. Luong, S.~D. {McNeil}, C.~S. Palshetkar, D.~C. Rafferty,
  C.~Simenel, K.~Stiefel,
 Phys. Rev. C
{\bf  91}, 041602 (2015).

\bibitem{wakhle2014}
A.~Wakhle, C.~Simenel, D.~J. Hinde, M.~Dasgupta, M.~Evers, D.~H. Luong,
  R.~du~Rietz, E.~Williams,
 Phys. Rev. Lett.
{\bf  113}, 182502 (2014).

\bibitem{adamian2003}
G.~G. Adamian, N.~V. Antonenko, W.~Scheid,
 Phys. Rev. C
{\bf  68}, 034601 (2003).

\bibitem{zagrebaev2007}
{Valery Zagrebaev}, {Walter Greiner},
 J. Phys. G
{\bf  34}, 2265 (2007).

\bibitem{aritomo2009}
Y.~Aritomo,
 Phys. Rev. C
{\bf  80}, 064604 (2009).

\bibitem{zhao2016}
Kai Zhao, Zhuxia Li, Yingxun Zhang, Ning Wang, Qingfeng Li, Caiwan Shen,
  Yongjia Wang, Xizhen Wu,
 Phys. Rev. C
{\bf  94}, 024601 (2016).

\bibitem{umar2015c}
A.~S. Umar, V.~E. Oberacker,
 Nucl. Phys. A
{\bf  944}, 238 (2015).

\bibitem{umar2016}
A.~S. Umar, V.~E. Oberacker, C.~Simenel,
 Phys. Rev. C
{\bf  94}, 024605 (2016).

\bibitem{prasad2016}
E.~Prasad, A.~Wakhle, D.~J. Hinde, E.~Williams, M.~Dasgupta, M.~Evers, D.~H.
  Luong, G.~Mohanto, C.~Simenel, K.~Vo-Phuoc,
 Phys. Rev. C
{\bf  93}, 024607 (2016).

\bibitem{feng2009a}
Zhao-Qing Feng, Gen-Ming Jin, Jun-Qing Li,
 Phys. Rev. C
{\bf  80}, 067601 (2009).

\bibitem{gupta2007b}
Raj~K. Gupta, S.~K. Patra, P.~D. Stevenson, Walter Greiner,
 Intl. J. of Mod. Phys. E
{\bf  16}, 1721 (2007).

\bibitem{sargsyan2009}
V.~V. Sargsyan, Z.~Kanokov, G.~G. Adamian, N.~V. Antonenko, W.~Scheid,
 Phys. Rev. C
{\bf  80}, 047603 (2009).

\bibitem{zhao2009}
Kai Zhao, Xizhen Wu, Zhuxia Li,
 Phys. Rev. C
{\bf  80}, 054607 (2009).

\bibitem{tian2008}
{Junlong Tian}, {Xizhen Wu}, {Kai Zhao}, {Yingxun Zhang}, {Zhuxia Li},
 Phys. Rev. C
{\bf  77}, 064603 (2008).

\bibitem{cusson1980}
R.~Y. Cusson, J.~A. Maruhn, H.~St\"ocker,
 Z. Phys. A
{\bf  294}, 257 (1980).

\bibitem{golabek2009}
{C\'edric Golabek}, {C\'edric Simenel},
 Phys. Rev. Lett.
{\bf  103}, 042701 (2009).

\bibitem{kedziora2010}
{David J. Kedziora}, {C\'edric Simenel},
 Phys. Rev. C
{\bf  81}, 044613 (2010).

\bibitem{simenel2012}
C\'edric Simenel,
 Eur. Phys. J. A
{\bf  48}, 152 (2012).

\bibitem{simenel2008}
{C\'edric Simenel}, {Benoit Avez},
 Intl. J. Mod. Phys. E
{\bf  17}, 31 (2008).

\bibitem{washiyama2008}
{Kouhei Washiyama}, {Denis Lacroix},
 Phys. Rev. C
{\bf  78}, 024610 (2008).

\bibitem{simenel2017}
C.~Simenel, A.~S. Umar, K.~Godbey, M.~Dasgupta, D.~J. Hinde,
 Phys. Rev. C
{\bf  95}, 031601 (2017).

\bibitem{umar2014a}
A.~S. Umar, C.~Simenel, V.~E. Oberacker,
 Phys. Rev. C
{\bf  89}, 034611 (2014).

\bibitem{vophuoc2016}
K.~Vo-Phuoc, C.~Simenel, E.~C. Simpson,
 Phys. Rev. C
{\bf  94}, 024612 (2016).

\bibitem{liang2016}
J.~F. Liang, J.~M. Allmond, C.~J. Gross, P.~E. Mueller, D.~Shapira, R.~L.
  Varner, M.~Dasgupta, D.~J. Hinde, C.~Simenel, E.~Williams, K.~{Vo--Phuoc},
  M.~L. Brown, I.~P. Carter, M.~Evers, D.~H. Luong, T.~Ebadi, A.~Wakhle,
 Phys. Rev. C
{\bf  94}, 024616 (2016).

\bibitem{godbey2017}
K.~Godbey, A.~S. Umar, C.~Simenel,
 Phys. Rev. C
{\bf  95}, 011601 (2017).

\bibitem{negele1982}
J.~W. Negele,
 Rev. Mod. Phys.
{\bf  54}, 913 (1982).

\bibitem{umar2015a}
A.~S. Umar, V.~E. Oberacker, C.~Simenel,
 Phys. Rev. C
{\bf  92}, 024621 (2015).

\bibitem{sekizawa2016}
Kazuyuki Sekizawa, Kazuhiro Yabana,
 Phys. Rev. C
{\bf  93}, 054616 (2016).

\bibitem{umar2006c}
A.~S. Umar, V.~E. Oberacker,
 Phys. Rev. C
{\bf  73}, 054607 (2006).

\bibitem{maruhn2014}
J.~A. Maruhn, {P.--G.} Reinhard, P.~D. Stevenson, A.~S. Umar,
 Comp. Phys. Comm.
{\bf  185}, 2195 (2014).

\bibitem{chabanat1998a}
E.~Chabanat, P.~Bonche, P.~Haensel, J.~Meyer, R.~Schaeffer,
 Nucl. Phys. A
{\bf  635}, 231 (1998).

\bibitem{kluepfel2009}
P.~Kl\"uepfel, {P.--G.} Reinhard, T.~J. B\"urvenich, J.~A. Maruhn,
 Phys. Rev. C
{\bf  79}, 034310 (2009).

\bibitem{kortelainen2010}
M.~Kortelainen, T.~Lesinski, J.~More, W.~Nazarewicz, J.~Sarich, N.~Schunck,
  M.~V. Stoitsov, S.~Wild,
 Phys. Rev. C
{\bf  82}, 024313 (2010).

\bibitem{umar2006b}
A.~S. Umar, V.~E. Oberacker,
 Phys. Rev. C
{\bf  74}, 021601 (2006).

\bibitem{umar2009a}
A.~S. Umar, V.~E. Oberacker, J.~A. Maruhn, {P.--G.} Reinhard,
 Phys. Rev. C
{\bf  80}, 041601 (2009).

\bibitem{umar2010a}
A.~S. Umar, V.~E. Oberacker, J.~A. Maruhn, P.-G. Reinhard,
 Phys. Rev. C
{\bf  81}, 064607 (2010).

\bibitem{simenel2004}
C.~Simenel, Ph. Chomaz, G.~{de France},
 Phys. Rev. Lett.
{\bf  93}, 102701 (2004).

\bibitem{umar2010c}
A.~S. Umar, J.~A. Maruhn, N.~Itagaki, V.~E. Oberacker,
 Phys. Rev. Lett.
{\bf  104}, 212503 (2010).

\bibitem{balian1992}
R.~Balian, M.~V\'en\'eroni,
 Ann. Phys.
{\bf  216}, 351 (1992).

\bibitem{lacroix2014}
Denis Lacroix, Sakir Ayik,
 Eur. Phys. J. A
{\bf  50}, 95 (2014).

\bibitem{tanimura2017}
Yusuke Tanimura, Denis Lacroix, Sakir Ayik,
 Phys. Rev. Lett.
{\bf  118}, 152501 (2017).

\bibitem{ayik2017}
S.~Ayik, B.~Yilmaz, O.~Yilmaz, A.~S. Umar, G.~Turan,
 Phys. Rev. C
{\bf  96}, 024611 (2017).

\bibitem{sekizawa2017}
Kazuyuki Sekizawa,
 Phys. Rev. C
{\bf  96}, 014615 (2017).

\bibitem{umar2017}
A.~S. Umar, C.~Simenel, W.~Ye,
 Phys. Rev. C
{\bf  96}, 024625 (2017).
\end{thebibliography}
\end{document}